\documentclass[aps,prd,twocolumn,superscriptaddress,showpacs,nofootnoteinbib]{revtex4}
\usepackage{natbib,hyperref,ifthen} 
\usepackage{graphicx}
\usepackage{dcolumn}
\usepackage{bm}
\usepackage{natbib}
\usepackage{multirow}
\usepackage{epsfig}
\usepackage{amsmath}

\newcommand{\beq}{\begin{equation}}
\newcommand{\eeq}{\end{equation}}
\newcommand{\beqa}{\begin{eqnarray}}
\newcommand{\eeqa}{\end{eqnarray}}

\newcommand{\be}{\begin{equation}}
\newcommand{\ee}{\end{equation}}

\newcommand{\rmd}{{\rm d}}
\renewcommand{\vec}[1]{{\bf #1}}

\newcommand{\fnl}{{f_{\rm NL}}}

\begin{document}

\title{Optimal weighting in $\fnl$ constraints from large scale
  structure in an idealised case}

\author{An\v{z}e Slosar} \affiliation{Berkeley Center for Cosmological
  Physics, Physics Department and Lawrence Berkeley National
  Laboratory,University of California, Berkeley California 94720, USA}
\affiliation{Faculty of Mathematics and Physics, University of
  Ljubljana, Slovenia}

\date{\today}

\begin{abstract}
  We consider the problem of optimal weighting of tracers of structure
  for the purpose of constraining the non-Gaussianity parameter
  $\fnl$. We work within the Fisher matrix formalism expanded around
  fiducial model with $\fnl=0$ and make several simplifying
  assumptions. By slicing a general sample into infinitely many
  samples with different biases, we derive the analytic expression for
  the relevant Fisher matrix element. We next consider weighting
  schemes that construct two effective samples from a single sample of
  tracers with a continuously varying bias. We show that a
  particularly simple ansatz for weighting functions can recover all
  information about $\fnl$ in the initial sample that is recoverable
  using a given bias observable and that simple division into two
  equal samples is considerably suboptimal when sampling of modes is
  good, but only marginally suboptimal in the limit where Poisson
  errors dominate.
\end{abstract}


\maketitle

\setcounter{footnote}{0}

\section{Introduction}
\label{sec:introduction}

The currently most attractive theory for the emergence of structure in
the Universe is inflation
\cite{Starobinsky:1979ty,1981PhRvD..23..347G,1982PhLB..108..389L,1982PhRvL..48.1220A}. It
is generically successful at diluting the primordial defects to
undetectable densities and predicts a nearly-flat universe with nearly
scale invariant spectrum of primordial fluctuations that are normally
distributed and extend to scales larger than horizon
\cite{1981JETPL..33..532M,1982PhLB..115..295H,1982PhRvL..49.1110G,1982PhLB..117..175S,1983PhRvD..28..679B}.
To understand details of the inflation, on must look at detailed
predictions of different models. Non-Gaussianity of the primordial
curvature perturbations, i.e small departures from the normal
distribution of fluctuations is one aspect in which models of
inflation differ.

Recently, non-Gaussianity of the local $\fnl$ type has received a
renewed attention.  This type of non-Gaussianity is characterised by a
quadratic correction to the potential
\citep{1990PhRvD..42.3936S,1994ApJ...430..447G,2000MNRAS.313..141V,2001PhRvD..63f3002K}:
\begin{equation}
  \Phi = \phi + \fnl \left( \phi^2 - \left<\phi^2 \right> \right),
\label{fnl}
\end{equation}
where $\phi$ is the primordial potential assumed to be a Gaussian
random field and $\fnl$ describes the amplitude of the correction
during the matter domination era.  There are two main reasons for this
renewed interest. First, there is a hint of a detection in the cosmic
microwave data \cite{2008PhRvL.100r1301Y} and several non-detections
\cite{2007JCAP...03..005C,2008arXiv0803.0547K,2008arXiv0802.3677H}.
Second, a new method for its detection has been recently proposed in
\cite{Dalal:2007cu}. This method uses biased tracers of structure for
which it can be shown that local-type of non-Gaussianity leads to a
very particular scale-dependence of the bias
\begin{equation}
  \Delta b = \fnl (b-1) u (k), 
\label{eq:4}
\end{equation}
where $\Delta b$ is the bias induced by non-Gaussianity, $b$ is the
tracer's intrinsic bias and $u$ is given by 
\begin{equation}
  u(k) = \frac{3 \delta_c \Omega_m H_0^2}{c^2 k^2 T(k) D(z)},
\label{eq:3}
\end{equation}
where $T(k)$ is the matter transfer function normalised to unity at
$k=0$, $D(z)$ is the growth function normalised to $(1+z)^{-1}$ in the
matter era, $\delta_c=1.68$ is the linear over-density at collapse for
the spherical collapse model and other symbols have their usual
meaning. Note that $\Delta b$ becomes significant only at large
scales, where non-linearities and scale-dependent bias are expected to
be small and therefore offers a surprisingly clean probe of
non-Gaussianity.  This equation has been re-derived, scrutinised and
better understood in the subsequent work
\cite{2008ApJ...677L..77M,2008arXiv0805.3580S,2008arXiv0806.1046A,2008arXiv0806.1061M}.

A first application of this method to the real data using a wide
variety of tracers of large scale has recently shown the promise of
this method \cite{2008arXiv0805.3580S,2008arXiv0806.1046A}. The
derived constraints are already competitive with those coming from the
cosmic microwave background.  In that work, the constraints were
derived by comparing the power spectrum of the distribution of tracers
with those predicted by the theory. At largest scales, where the
effect coming from the non-Gaussianity is the largest, the method
suffers from the sample variance. In other words, the finite number of
large-scale modes in any survey severely limits our ability to measure
the power spectrum.  Recently, Seljak has suggested a method of
circumventing this limitation \citep{2008arXiv0807.1770S}. This method
essentially considers two differently biased tracers that sample the
same volume. The ratio of amplitudes of a single mode for the two
tracers will give the ratio of the two biases $(b_1+\Delta
b_1(\fnl))/(b_2+\Delta b_2(\fnl))$, but the amplitude of the
primordial mode cancels out. One thus measures the auto correlation
power spectra of the two tracers in the same volume. By taking the
ratio of these two spectra, one can put a constraint on the value of
$\fnl$, which is independent on the primordial field and thus
unaffected by the sample variance. The biases $b_1$ and $b_2$ can be
derived from the amplitude of small scale fluctuations, where sampling
variance is not a problem and hence, one extremely well measured
large-scale mode is in principle enough to constrain $\fnl$.  A more
robust technique would be to assume nothing about the matter power
spectrum and derive limits on $\fnl$ from limits on the scale
dependence of ratio of $b+\Delta b$. This would protect measurements
of $\fnl$ from systematics arising from, for example,  massive neutrinos.

In practice, one rarely has two distinct samples with a well-defined
bias.  In this work, we extended the analysis by considering a single
tracer of the underlying field that spans a range of biases and
attempt to answer the question of how to optimally analyse such
tracer. The approach we take is to create two effective samples and
to optimally weight the tracer's constituents.

\section{Approach and limitations of this work}
\label{sec:appr-limit-this}
In this paper we assume that the Equations \eqref{eq:4} and
\eqref{eq:3} are exactly correct. These equations have initially been
derived using Press-Schecter\cite{1974ApJ...187..425P} and related
formalisms.  They have now been tested \emph{for dark matter halos}
against $N$-body simulations in two publications
\cite{2008arXiv0811.2748D, 2008arXiv0811.4176P} with somewhat
differing conclusions. This is an issue that will have to be settled
before further measurements of $\fnl$ are possible.

If luminous objects are used for constraining the $\fnl$ parameter, they must
sample the underlying population of halos randomly in the
sense that they must insensitive to any property of the halo that might be
correlated with the large-scale modes that induce the $\fnl$
dependence. While this is true for most objects, it is not
necessarily true for quasars as discussed in
\cite{2008JCAP...08..031S}, where the Equation \eqref{eq:4} was
generalised to $\Delta b = \fnl (b-p) u (k),$ with $p=1$ for random
halos and $p\sim 1.6$ for a population that is hosted by the recently
merged halos. This is, of course, a rather crude approximation, but it
illustrates a possible violation of the Equations \eqref{eq:4} and \eqref{eq:3}.

If this assumption holds then the stochasticity of each tracer will be
zero on scales much larger than the typical halo size. Stochasticity is
a measure of how well a given tracer of large scale structure samples
the underlying dark matter field in the Gaussian cosmologies. In this
work we consider a single tracer whose constituents have a range of
biases and zero stochasticity. In particular, each galaxy (or quasar
or some other tracer) has an associated bias $b$, so that a subset of
galaxies whose biases lie between $b$ and $b+\Delta b$ is a perfect
tracer of the underlying dark matter field with a constant bias
$b$. In other words, in Fourier space on scales of interest,
\begin{equation}
  \delta_g (\vec{k}) = b \delta (\vec{k}),
\end{equation}
where $\delta_g$ is the over-density of galaxies and $\delta$ is the
over-density of matter at some wave-vector $\vec{k}$. This assumption
can be checked by considering the quantity
$P_{12}(k)/\sqrt{P_{11}(k)P_{22}(k)}$, where $P_{12}$ is the
cross-correlation power spectrum for the two samples and
$P_{11}$/$P_{22}$ are the corresponding auto-correlation power
spectra. Recent cross-correlation studies using counts in cells have
shown that on large scales ($>10$ Mpc$/h$) the stochasticity is indeed
very small \cite{Swanson:2007kh}.  It is not fundamental limit, but it
means that improvement in signal to noise will stop beyond $\bar{n} =
(P(1-r^2))^{-1}$ \cite{2008arXiv0807.1770S}, where $r$ is the
cross-correlation coefficient. It is also possible the clever schemes
around this limitation might be constructed
\cite{2008arXiv0810.0273B}.

Moreover, we assume that the noise associated with the sparse sampling
of the underlying field can be described by a Poisson
statistics. While this sounds a very reasonable approximation, recent
work on $N$-body simulations indicate that the actual shot noise
properties might significantly deviate from Poisson statistics\footnote{Uro\v{s}
  Seljak, private communication.}.

Next we assume that there exist an observable that can be thought of
as a proxy for the individual galaxy's bias. In practice, this can be
the galaxy's luminosity, but in this paper we often operate with host
halo bias mass that allows us to connect our calculations with the
standard mass functions of the halo model. Since the bias is an
ill-defined quantity on a single object, it suffices that an ensemble
of galaxies with luminosity between $L$ and $L+\Delta L$ has a mean
bias $b(L)$. If $L$ is a ``noisy'' estimator for the galaxy's bias
then the range of biases obtainable from various slicing of the
original sample will be limited and hence any weighting based on $L$
will be suboptimal. In the limiting case when $L$ is a completely
random variable, any slicing would produce two samples of the same
bias and therefore it is impossible to put limits on $\fnl$ using
method of \cite{2008arXiv0807.1770S}. Our work derives the optimal
weighting within the possibilities offered by a given measurable
quantity and not optimal weighting in an absolute sense.

The purposed of this paper is to derive the optimal weighting subject
to limitations described above. However, it must be stressed that if
these conditions are not satisfied, the weighting will be suboptimal,
but it would not lead to biased results.  This is equivalent to the
inverse covariance weighting used in optimal quadratic estimators (see
e.g. \cite{1997PhRvD..55.5895T}). If wrong power spectrum is used to
create the covariance matrix, or if no inverse covariance weighting is
performed at all, the results are suboptimal and the error-bars are
larger then necessary, but results are not biased. Situation here is
similar: the method that we present here is trivial to implement on
real data, while the truly optimal weighting would require massive
numerical work. We therefore deem it a useful step towards decreasing
the error-bars on $\fnl$ constraints in future observational work.

The paper is organised as follows. In Section \ref{sec:fundamental-limit},
we consider slicing the sample into infinitely thin subsamples of
varying bias and derive an analytic expression for the maximum
signal-to-noise that can be obtained using a Fisher matrix
analysis. In the subsequent Section \ref{sec:optimal-weighting} we
consider how the sample can be weighted using two weighting functions
to get two effective samples.  We construct a weighting method whose
Fisher matrix element for $\fnl$ is the same as those of the optimal
analysis and is thus itself optimal. We show that simple methods of
dividing the sample into two can be considerably ineffective. Section
\ref{sec:comp-with-power} briefly compares our results with optimal
weights used in power spectrum determination. Final thoughts can be
found in the Conclusions.

\section{Information content in a tracer}
\label{sec:fundamental-limit}

Consider a tracer of mass that is composed of many individual objects
that have different biases with respect to the underlying density
field. For simplicity, let us assume that the variable that determines
an individual object's bias is its host halo mass, but note that in
general it can be any continuous variable that varies monotonically
with bias. The population is then characterised by $b(M)$, the average
bias of the objects with mass $M$ and the mass function $\rmd n/\rmd
M$, which is the number density of objects with mass between $M$ and
$M+\rmd M$. Let slice the total number of objects into $N$ samples of
different average bias. Each slice is centred around mass $M_i =
M_{\rm min} + (i-1/2) \Delta M$, where $\Delta M = (M_{\rm max}-M_{\rm
  min})/N$ and has bias $b_i = b(M_i)$ with number density of $n_i =
\rmd n / \rmd M (M_i) \Delta M$.  Following
\cite{2008arXiv0807.1770S}, we consider one Fourier mode of the
underlying density field. Its covariance matrix has the form

\begin{multline}
  C_{ij} = < \delta_i \delta_j > = \frac{1}{V} 
(b_i+(b_i-p)u\fnl) \\ \times (b_j+(b_j-p)u\fnl)P
+ \frac{\delta^{K}_{ij}}{n_iV}.
\label{eq:mat}
\end{multline}

Our ability to constrain $\fnl$ is determined by the Fisher matrix,
whose $\fnl$ elements are
\begin{equation}
  F_{\fnl \fnl } =  \frac{1}{2} {\rm Tr} \left[\vec{C}_{,\fnl} \vec{C}^{-1} \vec{C}_{,\fnl} \vec{C}^{-1} \right],
\label{eq:ff}
\end{equation}
evaluated at our fiducial model, which has $\fnl=0$. In that limit we
have
\begin{eqnarray}
  C_{ij} = \frac{1}{V} 
\left(\frac{\delta^{K}_{ij}}{n_i} + b_ib_jP\right) \label{eq:cij} \\
\left(C_{,\fnl}\right)_{ij} =  VPu\   (2 b_i b_j - pb_i -pb_j)
\end{eqnarray}

In Appendix \ref{ref:app} we show that the inverse of $C$ is given by
\newcommand{\db}{\left<\Delta b^2\right>}
\newcommand{\mb}{\left<b\right>} \newcommand{\mbb}{\left<b^2\right>}
\begin{equation}
  C^{-1}_{ij} = V \left (n_i \delta^K_{ij} - \frac{n_i n_j b_i b_j P}{1+\bar{n}P\mbb}\right),
\end{equation}
where we have replaced sums with the integrals and defined averages to be over the mass function:
\begin{eqnarray}
  \bar{n} = \int_{M_{\rm min}}^{M_{\rm max}} \frac{\rmd n}{\rmd M}
  \rmd M\\
  \left<b \right> = \frac{1}{\bar{n}} \int_{M_{\rm min}}^{M_{\rm max}} \frac{\rmd n}{\rmd M}
  b(M) \rmd M\\
  \left<b^2 \right> = \frac{1}{\bar{n}} \int_{M_{\rm min}}^{M_{\rm max}} \frac{\rmd n}{\rmd M}
  b^2(M) \rmd M
\end{eqnarray}
After some cumbersome, but straight-forward algebra, we arrive at
\begin{equation}
  F_{\fnl \fnl} = (u P \bar{n})^2 \left(C_0+C_1x+C_2x^2\right),
\label{eq:result}
\end{equation}
where
\begin{equation}
  x= \frac{P\bar{n}}{1+P\bar{n}\mbb}
\end{equation}
and
\begin{eqnarray}
C_0= 2\mbb^2 -4\mbb\mb p +\mb^2 p^2+\mbb p^2\\
\nonumber C_1= -4\mbb^3 +8\mbb^2\mb p -\mbb^2 p^2 \\
 -3\mbb\mb^2 p^2  \\
C_2=2\mbb^4-4\mbb^3\mb p+2\mbb^2\mb^2 p^2
\end{eqnarray}
This result encodes that maximum information that can be extracted
from a sample of objects.

\begin{figure}
  \centering

  \includegraphics[width=\linewidth]{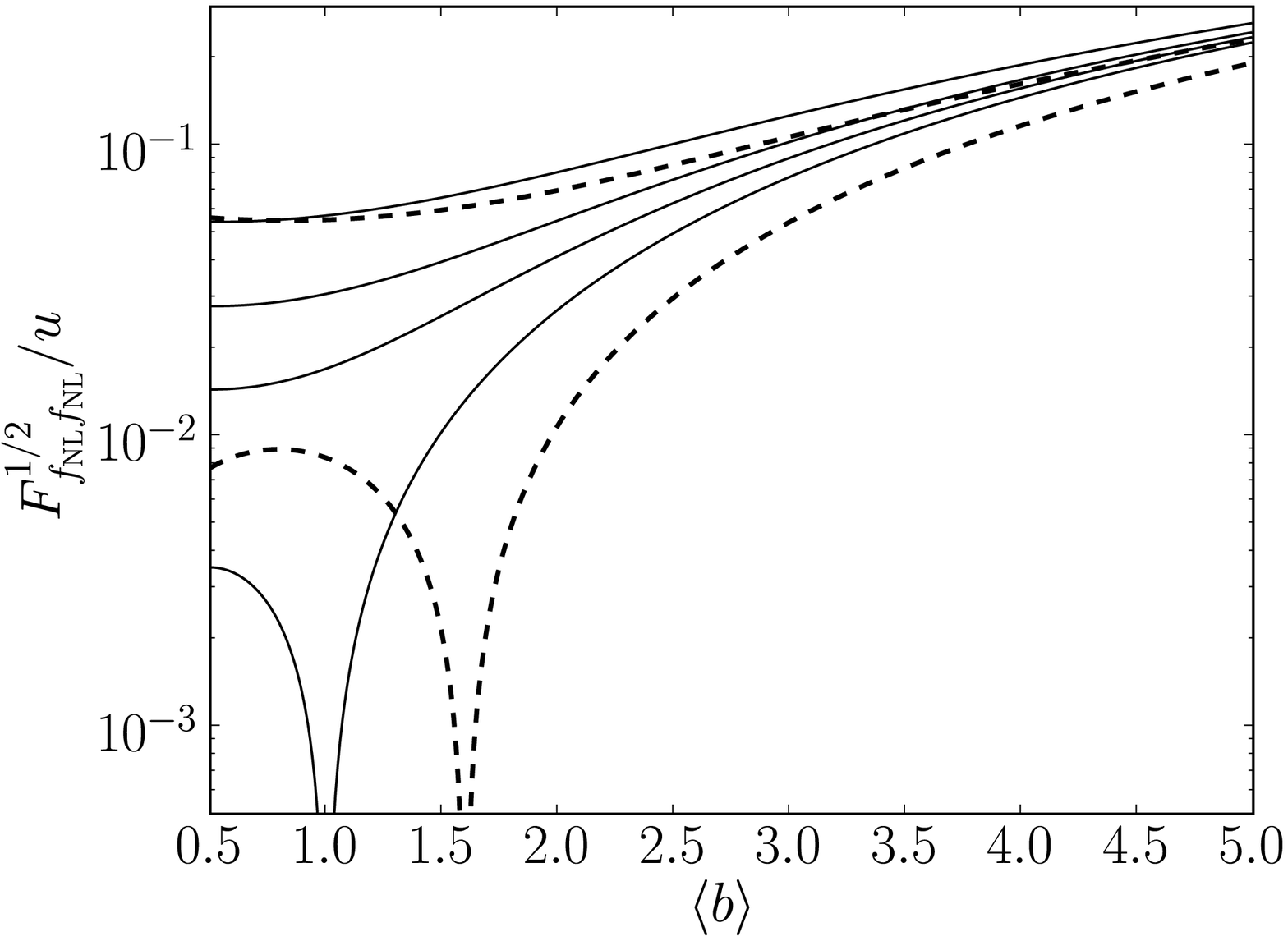} \\ 
  \includegraphics[width=\linewidth]{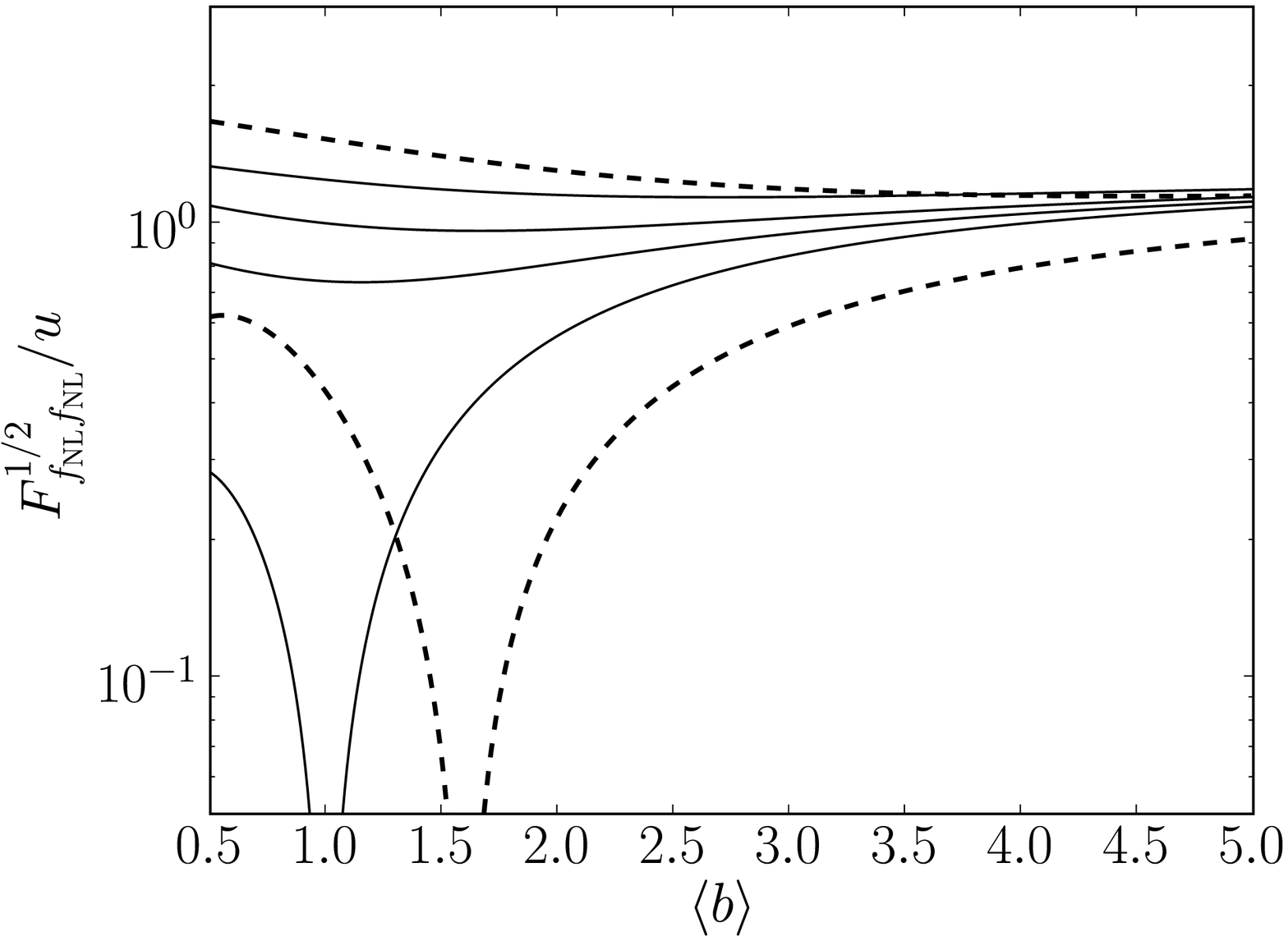} \\ 
  \includegraphics[width=\linewidth]{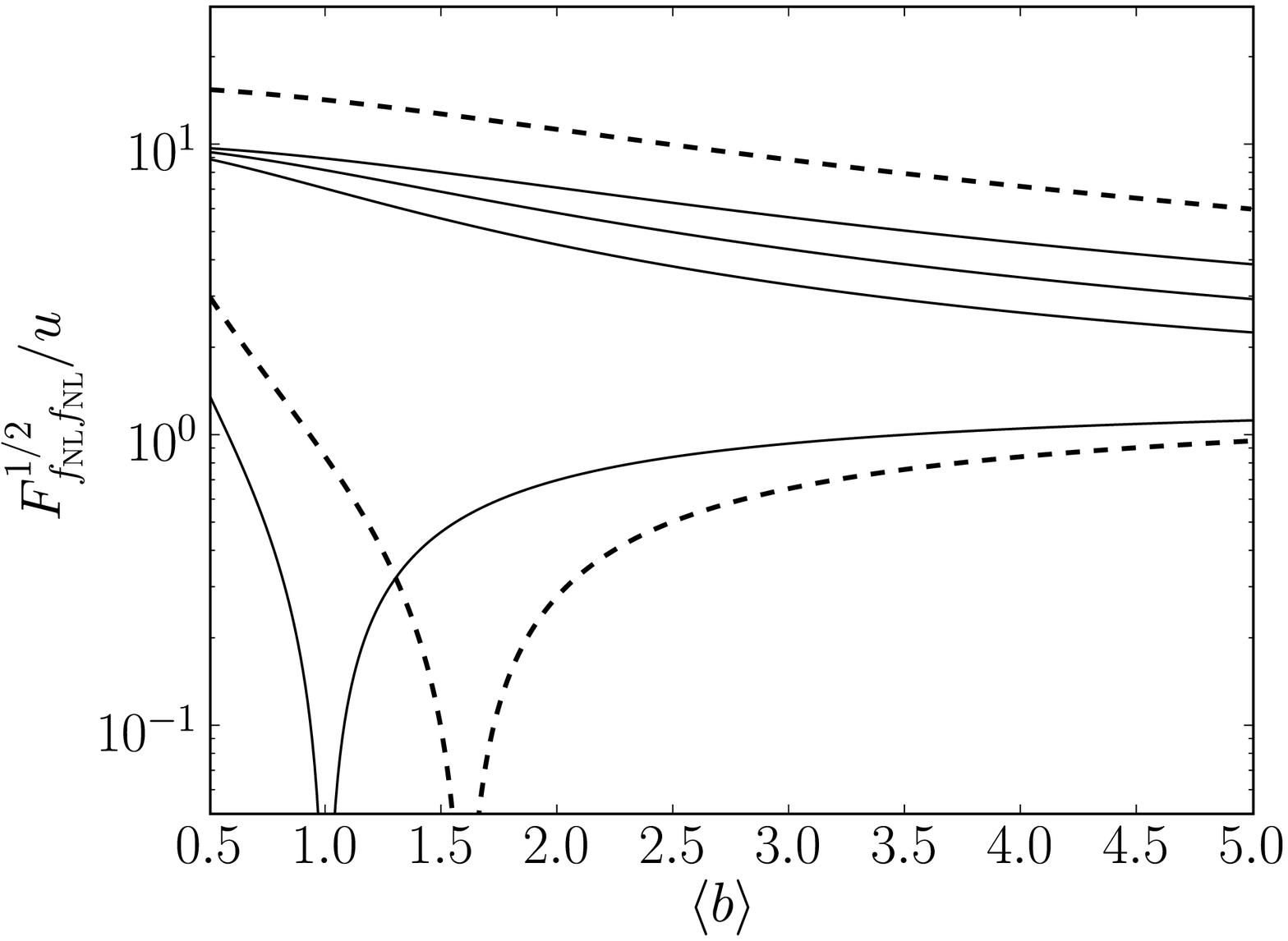} \\
  \caption{This figure shows scaling of $F_{\fnl \fnl}$ with $\mb$,
    $\db$ and $\bar{n}P$. Panels from the top to bottom correspond to
    values of $\bar{n}P$ of $10^{-2}$, $1$ and $10^2$, where $\bar{n}$
    is the tracer's number density and $P$ the underlying power
    spectrum. In each panel, thin solid lines correspond to values of
    $\db=0,1,2,4$ (bottom up) and $p=1$. Solid dashed lines are for
    $\db=0,4$ and $p=1.6$.}
  \label{fig:1}
\end{figure}

To get a better intuition about this formula, we define
\begin{equation}
  \left<\Delta b^2\right> = \left< b^2 \right> - \left<b\right>^2.
\end{equation}
In the Figure \ref{fig:1} we plot the functional shape for a couple of
values of $\mb$, $\db$ and $P\bar{n}$.

This figure deserves some discussion. As expected in the limit of
$\mbb=0$, the signal to noise drops to zero at $\mb=p$ and
monotonically increases with bias. In general, however,
this is not the case. When we are in the Poisson limit and sampling is
sparse, then it is still better to go with objects with the highest
bias. In the other limit, when sampling of the modes is very good, it
is better to have a bigger relative spread in the bias rather than
bias that is high in average. This is slightly counter-intuitive, but
remember that we assume here that each object has a known bias.  But
most importantly, when $\bar{n}P \sim 1$ the overall best signal to
noise is roughly independent of the mean bias, as long as we cover a
sizeable range of biases. 

This analysis corresponds to a single mode. For any realistic survey,
one needs to integrate across observed modes. The final error on
$\fnl$ is given by
\begin{equation}
  \sigma_{\fnl}^{-2} = \frac{V}{2 \pi^2}\int_{k_{\rm min}=\pi/V^{1/3}}^{\infty} F_{\fnl \fnl}
  k^2 \rmd k,
\label{eq:int}
\end{equation}
where $V$ is the volume of the survey and the pre-factors come from
the volume of a single mode in the $k$-space which equals $\pi^3/V$.

The result of the Equation  \eqref{eq:result} is the maximum information
that is in principle available for extraction. In practice, it is not
clear, how to extract this information - it would require a very fine
slicing by the bias with cross-correlation of each slice with every
other slice. In the next section we attempt a different approach - we
divide the sample into two different samples and adjusts the weighting
of the objects in the two samples so that the signal is maximised.

\section{Optimal weighting}
\label{sec:optimal-weighting}
Following the previous section,  we will consider weights that are
function of the halo mass $M$. In practice, we do not know the host halo
mass for individual objects, but one can equivalently use any proxy
for bias, such as luminosity. 

Let us therefore consider two weighting functions $\alpha(M)$ and
$\beta(M)$.  Any given object in the $\alpha$ sample counts as
$\alpha(M)$ objects. For example, when calculating the over-density in
a cell, we weight the objects by $\alpha(M)$:
\begin{equation}
  \delta = \frac{\sum_i \alpha(M_i)}{V_{\rm cell}N_\alpha}-1,
\end{equation}
where index $i$ runs over the halos in a given cell of volume $V_{\rm
  cell}$ and the mean weighted object density is given by
\begin{equation}
  N_\alpha = \int \alpha(M) \frac{\rmd n}{\rmd M} \rmd n
\end{equation}
and the same for the $\beta$ sample.

Using properties of the Poisson statistics, the effective bias and
corresponding Poisson error are given by
\begin{eqnarray}
\left(b_{\rm eff}\right)_\alpha  = \frac{1}{N_\alpha} \int \alpha(M) \frac{\rmd n}{\rmd M} b(M)
\rmd M,   \\
\left(\frac{1}{n_{\rm eff}}\right)_{\alpha\alpha}  = \frac{1}{N_\alpha^2} \int \alpha(M)^2 \frac{\rmd n}{\rmd M}
\rmd M
\end{eqnarray}
and an equivalent expression for the $\beta$ sample. An important
subtlety is, that if the weighting functions overlap, the cross term
also acquires a Poisson error, given by

\begin{equation}
\left(\frac{1}{n_{\rm eff}}\right)_{\alpha\beta}  = \frac{1}{N_\alpha N_\beta} \int
\alpha(M) \beta(M) \frac{\rmd n}{\rmd M} b(M)
\rmd M,
\end{equation}

For a two-sample case, the final error on the $\fnl$ can therefore be
calculated by combining Equations (\ref{eq:mat}) (with new Poisson
errors in the cross term), (\ref{eq:ff}) and (\ref{eq:int}).  Note
that results are independent of any multiplicative constant on
$\alpha$ or $\beta$.  However, one cannot assume that we are in the
Poisson limit and therefore the matrix inversions have to be done
without approximations. However, since we are discussing $2\times2$
matrices, this is not impossible.

Consider next the following form for weighting functions $\alpha$ and
$\beta$:
\begin{eqnarray}
  \alpha =  c_\alpha + b(M) \label{eq:1}\\
  \beta = c_\beta - b(M) \label{eq:2}
\end{eqnarray}

In this case, the relevant variables simplify to:
\begin{eqnarray}
  N_\alpha = \bar{n} \left(c_\alpha+\mb \right) \label{eq:xst}\\
  N_\beta = \bar{n} \left(c_\beta-\mb \right)\\
\left(b_{\rm eff}\right)_\alpha = \frac{c_\alpha\mb
  +\mbb}{c_\alpha+\mb}\\
\left(b_{\rm eff}\right)_\beta = \frac{c_\beta\mb
  -\mbb}{c_\beta-\mb}\\
\left(\frac{1}{n_{\rm eff}}\right)_{\alpha\alpha}  = \frac{1}{\bar{n}}
\frac{\mbb + 2\mb c_\alpha+c_\alpha^2}{(c_\alpha+\mb)^2}\\
\left(\frac{1}{n_{\rm eff}}\right)_{\beta\beta}  = \frac{1}{\bar{n}}
\frac{\mbb - 2\mb c_\beta+c_\beta^2}{(c_\beta-\mb)^2}\\
\left(\frac{1}{n_{\rm eff}}\right)_{\alpha\beta}  = \frac{1}{\bar{n}}
\frac{\left(-\mbb+\mb(c_\beta-c_\alpha)+c_\alpha
    c_\beta\right)}{(c_\alpha+\mb)(c_\beta-\mb)} \label{eq:xend}
\end{eqnarray}
We can now combine Equations \eqref{eq:xst}~--~\eqref{eq:xend} with
Equations \eqref{eq:mat} and \eqref{eq:ff} to obtain expression for
$F_{\fnl \fnl}$. This is a very cumbersome process that is best done
with the help of a mathematical computer package. The final result,
however reduces to the exactly the same expression as that of Equation
\eqref{eq:result}. This is a very interesting result. It shows that
any weighting that has the form of Equations
(\ref{eq:1})~--~(\ref{eq:2}) produces optimal sensitivity to the
$\fnl$.

In order to avoid dealing with nearly singular matrices, it is in
practice advantageous to have weighting functions that have as little
overlap as possible. We therefore propose the following form the 
weighting functions:
\begin{eqnarray}
\alpha(M) = \frac{b(M)-b_{\rm min}}{b_{\rm max}-b_{\rm min}}\\
\beta(M) = \frac{b_{\rm max}-b(M)}{b_{\rm max}-b_{\rm min}},
\end{eqnarray}
where $b_{\rm min}$ and $b_{\rm max}$ are the minimum and the maximum
value of bias in the range of interest. These optimal weighting
functions are the main result of this paper.

How does this compare to other weighting functions? The simplest case
would be to divide the sample into two disjoint samples with no
overlap:
\begin{eqnarray}
  \alpha(M) = H(M-M_{b})\\
  \beta(M) = 1-\alpha(M),
\end{eqnarray}
where $H(x)$ is the Heaviside step function and the barrier mass
$M_{b}$ is a free parameter. This is essentially equivalent to the
analysis of \cite{2008arXiv0807.1770S}, where presumably an absolute
magnitude cut is proposed to create two samples of a different bias.

We choose two possible values for $M_{b}$. First we consider $M_b$
such that the integral $\int \rmd n/\rmd M\, b(M) \rmd M$ is the same
for both samples. Second we use the $M_b$ that is such as to minimise
the overall $\sigma_\fnl$.

\begin{figure}
  \centering
  \includegraphics[width=\linewidth]{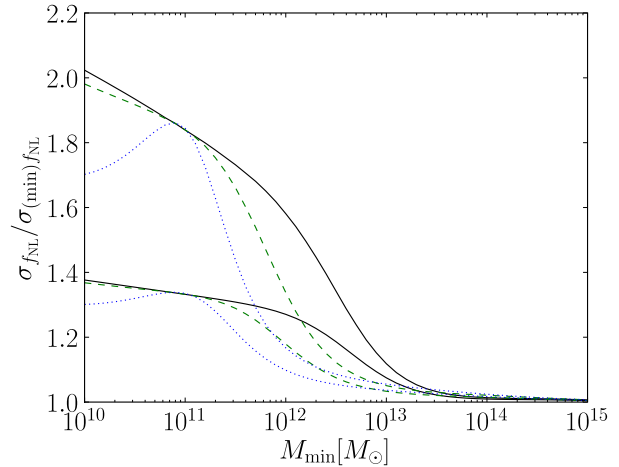}
  \caption{This figure shows the relative performance of the weighting
    methods that divide samples into two compared to optimal weighting
    for a model survey discussed in Section
    \ref{sec:fundamental-limit}. Top set of lines are for $q=b(M)$,
    while bottom are for the numerically determined optimal choice of
    $M_b$. Different line-styles represent density of objects to that
    of the halos: 1 (solid), 0.1 dashed and 0.01 (dotted). }
  \label{fig:buzo}
\end{figure}

In Figure \ref{fig:buzo} we plot how close the error on $\fnl$
approaches the theoretically minimal error obtained by optimal
weighting.  To plot this figure, we have assumed a fiducial flat
$\Lambda$CDM cosmology with matter density $\Omega_m=0.25$, spectral
index of primordial fluctuations of $0.96$ and normalisation in 8
Mpc/$h$ spheres of $\sigma_8=0.85$.  Moreover, we assumed a survey
centred at redshift $z=0.5$ with volume $V=1 ({\rm Gpc}/h)^3$ and
tracers with $p=1$ and used the mass function from the Sheth-Tormen
theory and bias from an extended Press-Schecter formalism
\citep{1999MNRAS.308..119S,2001MNRAS.323....1S}. The upper limit of
integration was set to $k_{\rm max}=0.05$ Mpc/$h$.

The x-axis of the plot is the minimal mass used for calculation of
$\mb$ and $\mbb$. The upper set of lines corresponds to a naive ansatz
of making $\int \rmd n/\rmd M\, b(M)\rmd M$ equal for both subsamples,
while the bottom set of lines for the best possible division that can
be obtained using two disjoint samples with no weighing. The dashed
and dotted lines show the dependence on the number-density of
objects. Note, that changing the number density affects both optimal
as well as suboptimal weighting and that we plot just the ratio of the
two error-bars, rather than the size of the error-bars themselves.

We see that the closer one is to the limit of well-sampled modes the
more important it is to use weighting, but that weighting does not
make any difference in the Poisson-limited sampling.  For the
particular survey parameter that we chose, we note that for $M_b$ set
by equal $\int \rmd n/\rmd M\, b(M) \rmd M$ for both samples, the
weighting function can be suboptimal to up to a factor of $\sim 2$ in
the limit of small $M_{\rm min}$, but that even the numerically
optimised $M_b$ can be significantly suboptimal.

The above results have to be take with a pinch of salt, since we have
used bias dependence on mass $b(M)$ and the corresponding mass
function $\rmd n/\rmd M$ rather than bias dependence on luminosity
$b(L)$ and the luminosity function $\rmd n/\rmd L$. The latter is more
closely related to the observations. Due to the scatter in luminosity
-- bias relation, the range of biases available by weighting by b(L)
is smaller and this will affect the results.

\section{Comparison with power spectrum weighting.}
\label{sec:comp-with-power}
Interestingly, the equations present in this paper are very close to
those that can be found in \cite{2004MNRAS.347..645P}. In that work,
authors find the optimal weighting for power spectrum determination
for a continuously biased tracer by generalising Feldman, Kaiser \&
Peacock \citep{1994ApJ...426...23F} approach. In fact, the equations
are very similar. Following exactly the same procedure as in Section
\ref{sec:fundamental-limit}, but for the power spectrum rather than
$\fnl$, one gets that in the limit of infinitely thin slicing
\begin{equation}
  F_{P P} = \left(\frac{\bar{n}\mbb}{1+P\bar{n}\mbb} \right)^2.
\end{equation}
By using the weighing function $\alpha(M) = b(M)$\footnote{We note that
  multiplicative factors on weighting function do not enter into the
  Fisher matrix analysis. One must nevertheless get them right to form
  an unbiased estimator.}, which is the optimal
weighting function found in \cite{2004MNRAS.347..645P} in our
notation, we can show that a single sample weighted with $\alpha(M)$
(i.e, the one-dimensional covariance matrix) again recovers the full
information $F_{P P}$. We therefore independently confirm the results
of \cite{2004MNRAS.347..645P}.

It is important to note that in this work, we have optimised for a
maximal $F_{\fnl \fnl}$ rather than a minimal $(F^{-1})_{\fnl
  \fnl}$. In other words, we calculate the weighting that maximises
our ability to constrain $\fnl$, assuming that other parameters, such
as $P$ are fixed and are presumably constrained from other probes,
such as cosmic microwave background.

\section{Conclusions}
\label{sec:conclusion}

In this paper we have analysed the problem of optimal weighting of
biased tracers of structure with the goal of extracting maximum
information about the non-Gaussianity parameter $\fnl$. We have
derived the minimum error on $\fnl$ by considering slices that are
infinitely thin in bias. We have shown that a simple weighting scheme
of Equations \eqref{eq:1} and \eqref{eq:2} obtains the same
constraining power. General division of the full sample into two
subsamples can be considerably sub-optimal even when mass at which the
samples are divided is carefully chosen.

The optimal weighting scheme of Equations \eqref{eq:1} and
\eqref{eq:2} is surprisingly simple. In fact, the product $P\bar{n}$
does not come into weighting at all - this is a lucky coincidence,
which allows us to use the same optimal weighting for every mode,
rather to optimize weighting around some fiducial wave-vector.

The result in this paper is subject to the assumptions outlined in the
Section \ref{sec:appr-limit-this} of this paper. If these assumptions
are violated, the weighing is sub-optimal, but probably nevertheless
beneficiary. Since any division into two samples by e.g. an absolute
magnitude cut requires some knowledge of bias, the implementation of
the scheme proposed in this paper is likely to be very simple.

How can this be put in practice? In this work we have used halo mass
$M$ as a proxy for the bias. However, our analysis is completely
general and one can replace the host halo mass with any variable that
is monotonically linked to the bias. For example, one could take
luminous red galaxies (LRGs) and determine their bias by splitting the
entire sample into several subsamples in different luminosity bins and
the constrain a smooth function $b(L)$, which describes the variation
of galaxy bias with its luminosity, using modes which are not affected
by the $\fnl$. One would next construct two effective samples by
optimally weighting the original sample using Equations \eqref{eq:1}
and \eqref{eq:2} and replacing $b(M)$ with $b(L)$. In the next step,
auto and cross-correlation power spectra of these two samples should
be calculated, taking into account the Poisson error correlation
between the two. At this step, one can use the cross-correlation
spectra to check for the amount of stochasticity, which has been
assumed to be negligible in this work. Finally, $\fnl$ should be
constrained using these power spectra as input.

\section*{Acknowledgements}

Numerical codes used in preparation of this paper used the mass
functions prepared using code by Darren Reed \cite{Reed:2006rw}.
Author thanks Will Percival for pointing out analogies with optimal
weighting of biased tracers for power spectrum estimation and
acknowledges useful discussions with Uro\v{s} Seljak.  This work is
supported by the inaugural BCCP Fellowship.

\appendix

\section{Inversion of $C$ matrix}
\label{ref:app}
We can rewrite Equation (\ref{eq:cij}) as 
\begin{equation}
  \vec{C} = \vec{N} \left( \vec{I}+\vec{E}\right),
\end{equation}
where $N_{ij}=\delta^{K}_{ij}V n_i^{-1}$, $\vec{I}$ is the identity
matrix and $E_{ij} = n_i b_i b_j P$. The inverse of $\vec{C}$ can then
formally be written as an infinite series
\begin{equation}
  \vec{C}^{-1} = \left(\vec{I}-\vec{E}+\vec{E}^2-\vec{E}^3 \ldots
  \right) \vec{N}^{-1}. \label{eq:tosim}
\end{equation}
We note that the product 
\begin{multline}
  \left( \vec{E}^2 \right)_{ij} = \sum_k n_i b_i b_k P n_k b_k b_j P =
  \\ E_{ij} \left(\sum_k n_k b_k^2 P \right) = \vec{E} \mbb
  \left(\bar{n}P \right),
\end{multline}
and so we can rewrite the inverse of $\vec{I}+\vec{E}$ as
\begin{multline}
  \left(\vec{I}+\vec{E}\right)^{-1} = \vec{I} - \vec{E} \left (1-\mbb
  \left(\bar{n}P \right) + \mbb^2 \left(\bar{n}P \right)^2 \ldots\right)\\
=\vec{I} - \vec{E}\frac{1}{1+\mbb  \left(\bar{n}P \right)}
\end{multline}
Since $\vec{N}$ is diagonal and hence trivial to invert, the Equation
(\ref{eq:tosim}) simplifies to 
\begin{equation}
    C^{-1}_{ij} = V \left (n_i \delta^K_{ij} - \frac{n_i n_j b_i b_j P }{1+\bar{n}P\mbb}\right)
\end{equation}

\bibliographystyle{arxiv}
\bibliography{cosmo,cosmo_preprints,fnl}

\end{document}